\documentclass[%
reprint,
superscriptaddress,
showpacs,
amsmath,amssymb,
aps,
pra,
]{revtex4-1}

\usepackage{amsmath}
\usepackage{graphicx}
\usepackage{braket}
\usepackage{amssymb}
\usepackage{float}
\usepackage{color}
\usepackage[normalem]{ulem}
\usepackage[colorlinks=true,linkcolor=blue,anchorcolor=blue,citecolor=blue,urlcolor=blue]{hyperref}
\begin{document}
\preprint{APS/123-QED}

\title{Nondipole Contributions to Attosecond Chiral Photoionization Asymmetries}

\author{Zheming Zhou}
\thanks{Equal contribution}
\affiliation{School of Physics and Wuhan National Laboratory for Optoelectronics, Huazhong University of Science and Technology, Wuhan 430074, China}
\author{Yang Li}
\thanks{Equal contribution}
\email{liyang22@sjtu.edu.cn}
\affiliation{State Key Laboratory of Dark Matter Physics, Key Laboratory for Laser Plasmas (Ministry of Education) and School of Physics and Astronomy, Collaborative Innovation Center for IFSA (CICIFSA), Shanghai Jiao Tong University, Shanghai 200240, China}
\author{Yueming Zhou}
\email{zhouymhust@hust.edu.cn}
\affiliation{School of Physics and Wuhan National Laboratory for Optoelectronics, Huazhong University of Science and Technology, Wuhan 430074, China}
\author{Peixiang Lu}
\affiliation{School of Physics and Wuhan National Laboratory for Optoelectronics, Huazhong University of Science and Technology, Wuhan 430074, China}



\begin{abstract}

Photoelectron circular dichroism (PECD) reads molecular chirality from forward-backward asymmetries in photoelectron emission, but the same observable can also contain non-dipole contributions from photon momentum transfer. Here we show that such contributions can reshape attosecond PECD measurements in both one- and two-photon ionization of chiral molecules. Calculations beyond the dipole approximation, interpreted with perturbation theory, reveal that non-dipole effects modify not only the magnitude but also the phase of the emitted electron wave packet. In two-photon interferometry, pathway interference amplifies the non-dipole response and can reverse the apparent chiral asymmetry. We further identify a practical separation principle: the non-dipole component is insensitive to enantiomeric handedness and can therefore be obtained from a racemic mixture. Subtracting this background isolates the purely chirality-induced asymmetry, enabling more accurate measurements of chiral electron dynamics.


\end{abstract}

\maketitle

\section{Introduction}
Chirality is a fundamental geometric property of matter and plays an essential role across chemistry, physics, and the life sciences. When chiral molecules are ionized by circularly polarized light, the emitted photoelectrons can exhibit a forward--backward asymmetry (FBA) along the light propagation direction that survives orientational averaging~\cite{PhysRevA.14.359}. This helicity-dependent and enantiosensitive asymmetry, known as photoelectron circular dichroism (PECD)~\cite{D4CP03770G}, is an electric-dipole effect that is typically much stronger than conventional circular dichroism~\cite{https://doi.org/10.1002/bbpc.19950990225}. PECD has therefore become a powerful observable for probing chiral photoionization dynamics and has been widely explored in one-photon ionization~\cite{10.1063/1.480581,PhysRevLett.86.1187,PhysRevA.98.063428,PhysRevLett.127.103201,PhysRevResearch.7.L012047}, strong-field ionization~\cite{PhysRevA.89.053406,Beaulieu_2016,doi:10.1126/science.aao5624,Comby2018,PhysRevLett.121.253201,PhysRevResearch.1.033045,PhysRevX.11.041056,PhysRevLett.126.083201,Rozen_2021,PhysRevLett.129.233201,PhysRevLett.129.243201,PhysRevResearch.6.043176,PhysRevA.110.013103,gcxt-18gk}, and pump--probe spectroscopy~\cite{Beaulieu2018,doi:10.1126/sciadv.abq2811,doi:10.1126/sciadv.ade0311,PhysRevX.13.011044,Wanie2024,PhysRevA.109.012810}. Most studies of chiral photoionization, however, have been formulated within the dipole approximation, in which the magnetic component of the light field and the spatial variation of the electric field are neglected. These non-dipole effects are known to generate propagation-direction asymmetries in photoelectron emission from atoms~\cite{PhysRevLett.106.193002,PhysRevLett.113.243001,PhysRevLett.113.263005,PhysRevLett.118.163203,Ilchen2018,Hartung2019,PhysRevLett.124.043201,PhysRevLett.125.073202,doi:10.1126/sciadv.abn7386,Liang2024,Mao2025,55pk-rxzg,doi:10.34133/ultrafastscience.0160} and molecules~\cite{doi:10.1126/science.abb9318,PhysRevLett.124.233201,PhysRevLett.129.253201}. Quantifying this competing source of asymmetry is therefore essential for precision measurements of chiral photoionization dynamics.

The first theoretical study of non-dipole PECD in chiral molecules considered inner-shell ionization~\cite{x18h-rn31}, where heavy atomic elements can enhance the effect. In contrast, the role of non-dipole interactions in outer-shell photoionization, which underlies most PECD measurements, remains largely unexplored. This question becomes especially relevant in attosecond interferometric measurements. Among them, reconstruction of attosecond beating by interference of two-photon transitions (RABBITT)~\cite{doi:10.1126/science.1059413} is a central technique for resolving electron dynamics on attosecond time scales and has been broadly applied in atomic and molecular systems~\cite{PhysRevLett.106.143002,PhysRevLett.117.093001,Cattaneo2018,PhysRevLett.123.133201,Han2023,PhysRevLett.131.203201,Loriot2024,doi:10.1126/sciadv.adj2629,PhysRevA.110.023109,jzfm-8vhj,z3k5-wqg9}. With recent advances in circularly polarized attosecond extreme-ultraviolet sources~\cite{Ferre2015,Huang2018,Han:23}, RABBITT has been extended to outer-shell ionization of chiral molecules~\cite{Han2025,8f7h-6nfc,10.1117/1.AP.8.1.015001}. In this configuration, the interference of two-photon pathways provides access to both the amplitude and phase of the emitted chiral electron wave packet, making it highly significant for probing and coherently controlling chiral electronic dynamics.

Here we investigate how non-dipole effects modify attosecond photoionization asymmetries in chiral molecular RABBITT measurements. Calculations beyond the dipole approximation, combined with second-order perturbation analysis, show that the measured FBA contains intertwined chiral and non-dipole contributions in both one- and two-photon ionization. The non-dipole contribution drives photoelectron emission preferentially along the light propagation direction and becomes stronger at higher electron energies, where it can reverse the sign of the measured FBA. Two-pathway interference amplifies this effect, moving the sign reversal to lower energies. Crucially, the non-dipole contribution is nearly independent of enantiomer, which allows the measured FBA to be decomposed into chiral and non-dipole parts. A racemic mixture can therefore provide an experimental calibration of the non-dipole background. We further show that non-dipole interactions produce few-attosecond forward-backward shifts in the RABBITT delay, which obey the same enantiomer-independent behavior. These results establish a practical route to separate non-dipole and chiral dynamics in high-precision attosecond measurements.

\section{Results and discussion}

\begin{figure}[t]
	\includegraphics[width=\columnwidth]{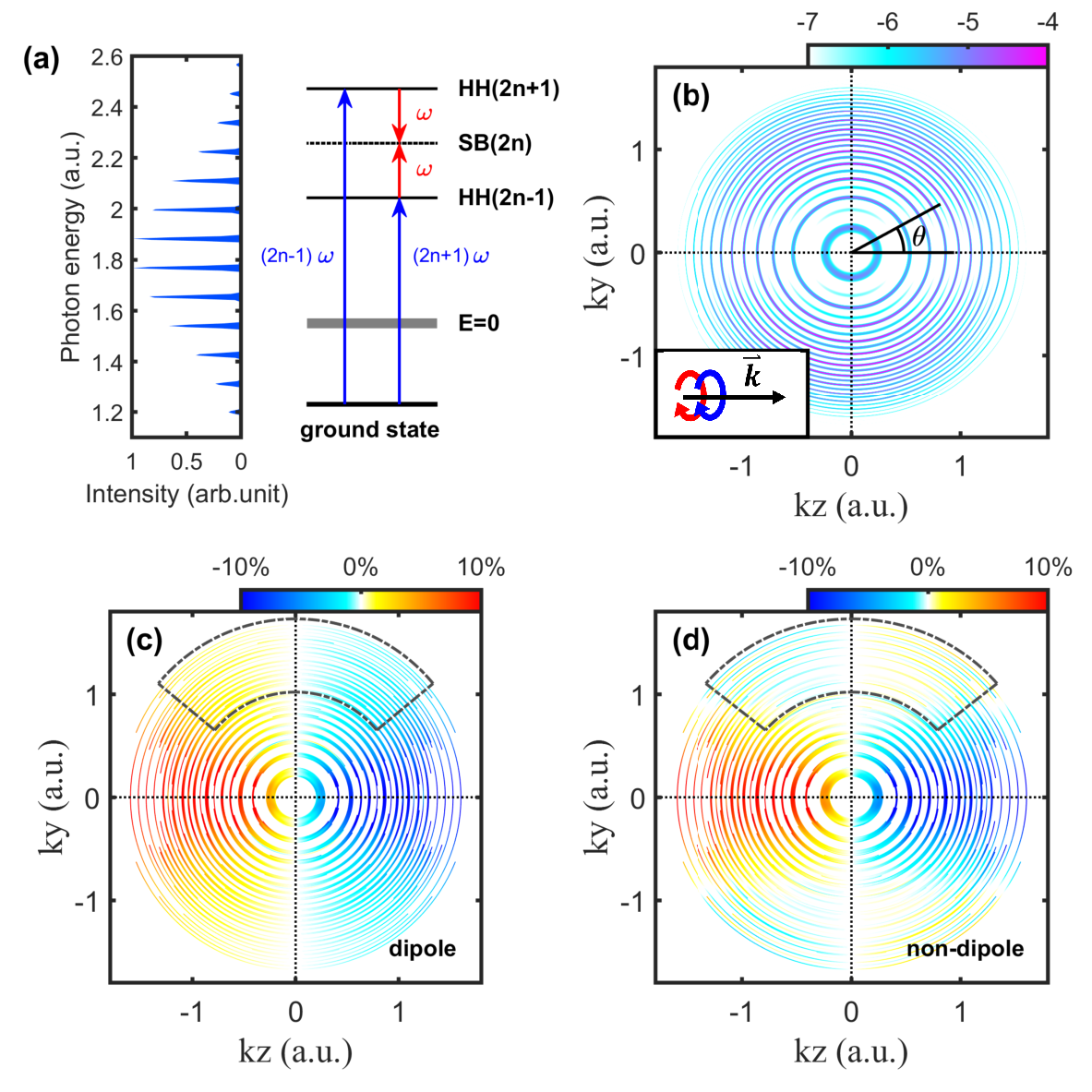}
	\caption{\label{F1}
         (a) Circularly polarized RABBITT scheme for chiral molecules and normalized spectrum of the XUV attosecond pulse train. (b) Photoelectron momentum distribution for enantiomer (+) driven by co-propagating counter-rotating circularly polarized fields. The peak intensities of the XUV and IR fields are $1\times10^{12}\,\mathrm{W/cm^{2}}$ and $1\times10^{11}\,\mathrm{W/cm^{2}}$, respectively, and the IR wavelength is 800 nm. The spectrum is plotted on a logarithmic scale, and the polar angle $\theta$ is measured with respect to the propagation axis. (c) FBA of the photoelectron momentum distribution in the dipole approximation. (d) FBA with non-dipole effects included.
	}
\end{figure}

We first consider the RABBITT scheme shown in Fig.~\ref{F1}(a). A randomly oriented ensemble of chiral molecules is ionized by a circularly polarized XUV attosecond pulse train propagating along the $z$ axis. The XUV field, which contains odd-order harmonics, drives outer-shell electrons into a set of main peaks. A weak co-propagating counter-rotating circularly polarized infrared (IR) field then couples adjacent main peaks and generates sidebands through two-photon interference. It has been shown that this interference could enhance the chirality-induced FBA signal~\cite{8f7h-6nfc} and enable photoionization time delays to be extracted from the sideband fringes~\cite{Han2025,10.1117/1.AP.8.1.015001}. We calculate the photoelectron momentum distributions by solving the TDSE of a model chiral molecule. Details of the TDSE propagation and molecular-orientation averaging procedure are provided in the Supplementary Material. Figure~\ref{F1}(b) shows the resulting momentum distribution in the $yz$ plane. In general, within the framework of second-order perturbation theory, the photoelectron momentum distribution can be expanded as
\begin{equation}
    \begin{aligned}
        I(k,\theta,\varphi)=\sum_{lm}\beta_{lm}(k)Y_{lm}(\theta,\varphi).
    \end{aligned}
    \label{E10}
\end{equation}
The allowed partial waves differ between the dipole approximation and the non-dipole calculation.  In the dipole approximation, the main peaks satisfy  $0\le l\le2$ with $m=0$, whereas the sidebands satisfy $0\le l\le4$ with $m=0,\pm2$. Including non-dipole terms extends these ranges to $0\le l\le3$ with $m=0$ for the main peaks and $0\le l\le6$ for the sidebands with $m=0,\pm2$ (see Supplementary Information for details). Figures~\ref{F1}(c) and \ref{F1}(d) compare the FBA in the dipole approximation and in the non-dipole calculation. We define
\begin{equation}
\mathrm{FBA}(k,\theta,\varphi)=2\frac{I(k,\theta,\varphi)-I(k,\pi-\theta,\varphi)}{I(k,\theta,\varphi)+I(k,\pi-\theta,\varphi)},
\end{equation}
which, using Eq.~\eqref{E10}, can be written as
\begin{equation}
    \begin{aligned}
        \mathrm{FBA}(k,\theta,\varphi)=2\frac{\sum_{l=\text{odd}}\beta_{lm}(k)Y_{lm}(\theta,\varphi)}{\sum_{l=\text{even}}\beta_{lm}(k)Y_{lm}(\theta,\varphi)}.
    \end{aligned}
    \label{E11}
\end{equation}
The sector marked in Fig.~\ref{F1} changes sign when non-dipole effects are included, demonstrating that photon-momentum transfer can strongly reshape the high-energy FBA. Because experiments cannot simply switch off non-dipole couplings, this contribution must be quantified before PECD-based observables can be interpreted as purely chiral effects. We therefore analyse the main peaks, which correspond predominantly to one-photon ionization, and the SBs, which arise from two-photon ionization, separately.

\begin{figure}[t]
    \includegraphics[width=\columnwidth]{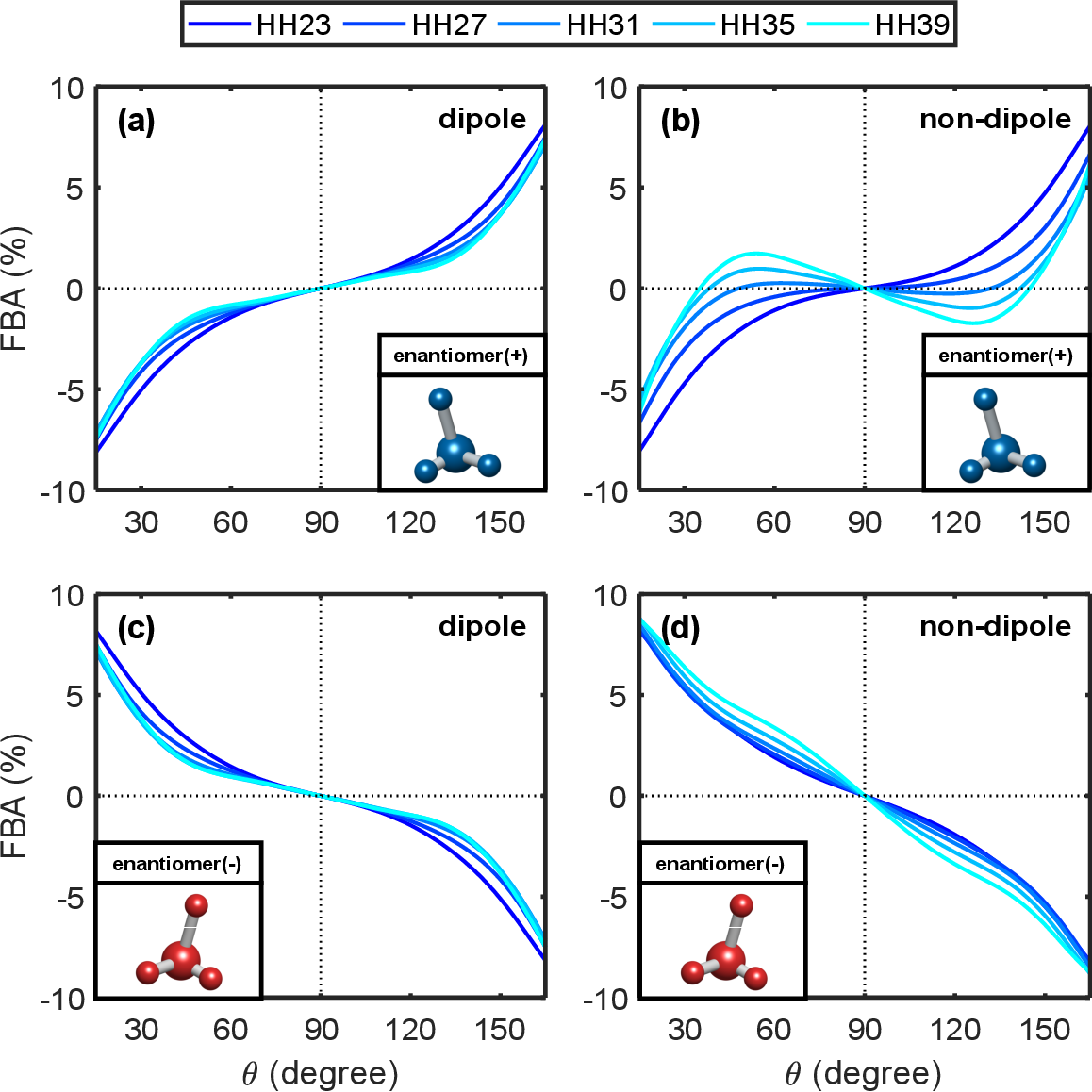}
    \caption{\label{F2}
        Angle-resolved FBA of the main peaks for enantiomer (+) (a) in the dipole approximation and (b) with non-dipole effects included. (c),(d) Same as (a),(b), but for enantiomer (-). The polar angle $\theta$ is measured with respect to the propagation axis.
        }
\end{figure}

Because the IR field is weak, its effect on the main peaks can be neglected. The main peaks can therefore be treated as one-photon ionization signals by the XUV field. To isolate non-dipole effects in outer-shell ionization, we calculated the angle-resolved FBA for both enantiomers under a counter-rotating polarization configuration, first in the dipole approximation and then with non-dipole interactions included. Figure~\ref{F2} shows the angle-resolved FBA for the five main peaks, ordered from low to high electron energy, for both enantiomers. In the dipole approximation, the two enantiomers FBAs with opposite signs and strong polar-angle dependence [Figs.~\ref{F2}(a) and \ref{F2}(c)]. The signal is largest near the propagation axis and is strongest at low electron energy, where the chiral scattering interaction is most pronounced; its magnitude reaches about 8\%. Including non-dipole terms changes the main-peak FBAs markedly. For enantiomer (+), a sign reversal appears beginning at the 31st-order main peak (19.45 eV) and strengthens at higher energies [Fig.~\ref{F2}(b)]. For enantiomer (-), the FBA is enhanced relative to the dipole result, especially in the high-energy region [Fig.~\ref{F2}(d)].

\begin{figure}[t]
    \includegraphics[width=\columnwidth]{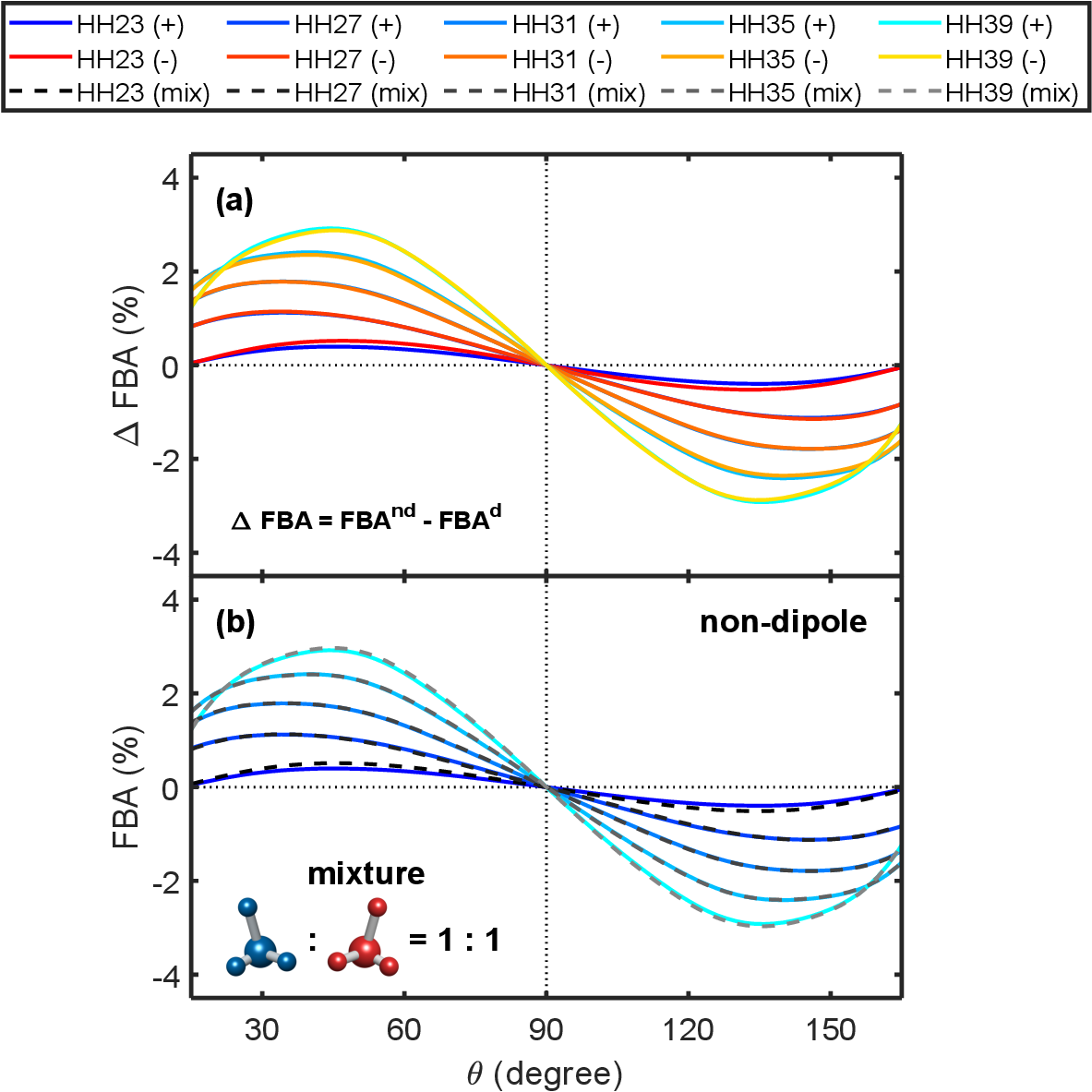}
    \caption{\label{F3}
        (a) Difference between the FBA including non-dipole effects and the FBA in the dipole approximation. Results for both enantiomers are shown. (b) FBA of the racemic mixture for each main peak (dashed lines), compared with $\Delta\mathrm{FBA}$ for enantiomer (+).
        }
\end{figure}

\begin{figure*}[t]
    \includegraphics[width=\textwidth]{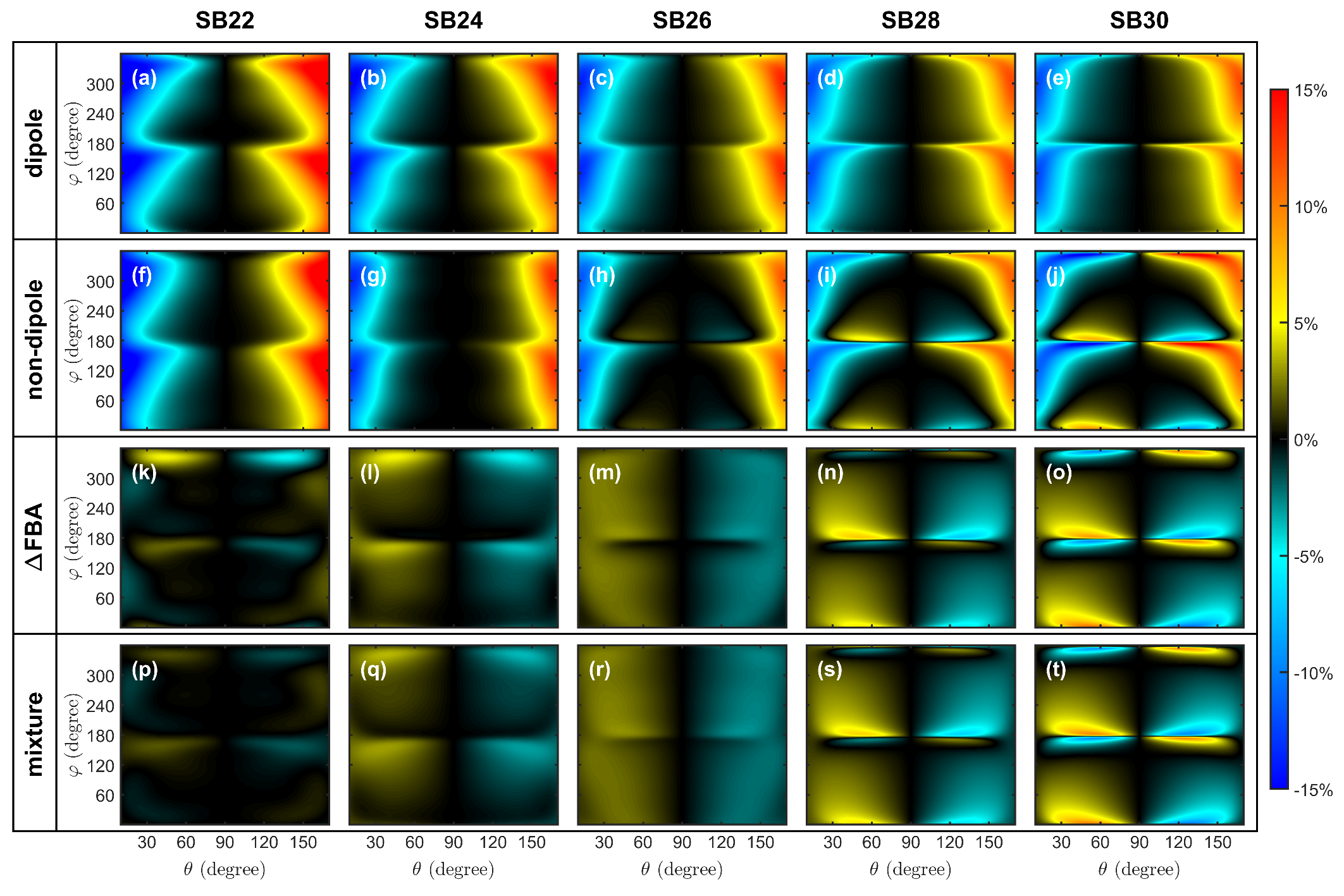}
    \caption{\label{F4}
        (a)--(e) Angle-resolved FBA for different sidebands in the dipole approximation. (f)--(j) Same as (a)--(e), but with non-dipole effects included. (k)--(o) Difference $\Delta\mathrm{FBA}$ between the non-dipole and dipole results. (p)--(t) Angle-resolved FBA for the racemic mixture. Here $\theta=\arccos{(k_{z}/\sqrt{k_{x}^{2}+k_{y}^{2}+k_{z}^{2}})}$ and $\varphi=\arg(k_{x}+ik_{y})$. Panels (a)--(o) correspond to enantiomer (+).
        }
\end{figure*}

To quantify the non-dipole contribution, we calculated the difference between the non-dipole and dipole FBAs, denoted $\Delta\mathrm{FBA}$. As shown in Fig.~\ref{F3}(a), the resulting $\Delta\mathrm{FBA}$ is centrosymmetric with respect to the origin: it is positive for $\theta<90^{\circ}$ and negative for $\theta>90^{\circ}$. Its magnitude grows away from the propagation axis ($\theta=0^{\circ}$) and then decreases as the emission approaches the polarization plane ($\theta=90^{\circ}$). This behaviour shows that non-dipole interactions favour photoelectron emission along the light propagation direction, with the strongest effect away from both the propagation axis and the polarization plane. Physically, this trend reflects photon-momentum transfer to the emitted electron during ionization. Most importantly, the two enantiomers yield nearly identical $\Delta\mathrm{FBA}$ curves. This near overlap behavior shows that the leading non-dipole contribution is enantiomer-insensitive. The odd-parity anisotropy parameters in Eq.~\eqref{E10} can therefore be separated, to leading order, into non-dipole and chiral contributions, $\beta_{lm}=\beta_{lm}^{(\text{nd})}\pm\beta_{lm}^{(\text{chiral})}$. Here $\beta_{lm}^{(\text{nd})}$ is the non-dipole contribution, whereas $\beta_{lm}^{(\text{chiral})}$ is the chiral contribution and changes sign when the enantiomer is reversed. Equation~\eqref{E11} then gives
\begin{equation}
    \begin{aligned}
        \mathrm{FBA}=\underbrace{2\frac{\sum_{l=\text{odd}}\beta_{lm}^{(\text{nd})}Y_{lm}}{\sum_{l=\text{even}}\beta_{lm}Y_{lm}}}_{\Delta\mathrm{FBA}}
        +\underbrace{2\frac{\sum_{l=\text{odd}}\beta_{lm}^{(\text{chiral})}Y_{lm}}{\sum_{l=\text{even}}\beta_{lm}Y_{lm}}}_{\mathrm{FBA}^{(\text{chiral})}},
    \end{aligned}
    \label{E12}
\end{equation}
which allows the measured FBA to be decomposed into a non-dipole contribution, $\Delta\mathrm{FBA}$, and a chiral contribution, $\mathrm{FBA}^{(\text{chiral})}$. Then, if the chiral-molecule ensemble is prepared as a racemic mixture, it cancels the chiral-geometry contribution and directly measures the non-dipole part of the FBA in chiral-molecule ionization, providing a direct experimental calibration of the non-dipole background.

We tested this prediction by calculating the FBA of a racemic mixture and comparing it with $\Delta\mathrm{FBA}$ for enantiomer (+) [Fig.~\ref{F3}(b)]. The two curves agree closely, giving $\mathrm{FBA}^{(\text{mix})}=\Delta\mathrm{FBA}$. Thus, the racemic mixture provides an experimental reference for the non-dipole contribution. Once this reference is measured, the purely chiral part can be obtained from $\mathrm{FBA}^{(\text{chiral})}=\mathrm{FBA}-\mathrm{FBA}^{(\text{mix})}$. This subtraction protocol provides a direct way to disentangle non-dipole and chiral responses in single-photon PECD measurements.

We next consider the SBs, where two-photon pathways interfere. Such interference is widely used in chiral RABBITT to enhance PECD and coherently control chiral electronic dynamics. It is therefore essential to understand how non-dipole effects enter the interferometric signal. Because the sideband distribution depends on both the polar angle $\theta$ and the azimuthal angle $\varphi$, we analysed the angle-resolved FBA for different SB orders. Figures~\ref{F4}(a)--\ref{F4}(e) show the dipole results for enantiomer (+). The FBA is strongest near the propagation axis and oscillates with $\varphi$. As the SB order increases, the signal gradually weakens, as in the one-photon case. When non-dipole effects are included [Figs.~\ref{F4}(f)--\ref{F4}(j)], sign reversals appear already at sideband 26 (11.70 eV) and become more prominent with increasing SB energy. Compared with one-photon ionization, two-pathway interference amplifies the non-dipole contribution and shifts the sign reversal to lower electron energy. Non-dipole effects are therefore more pronounced and non-negligible in interferometric measurements of chiral photoionization.

\begin{figure}[t]
    \includegraphics[width=\columnwidth]{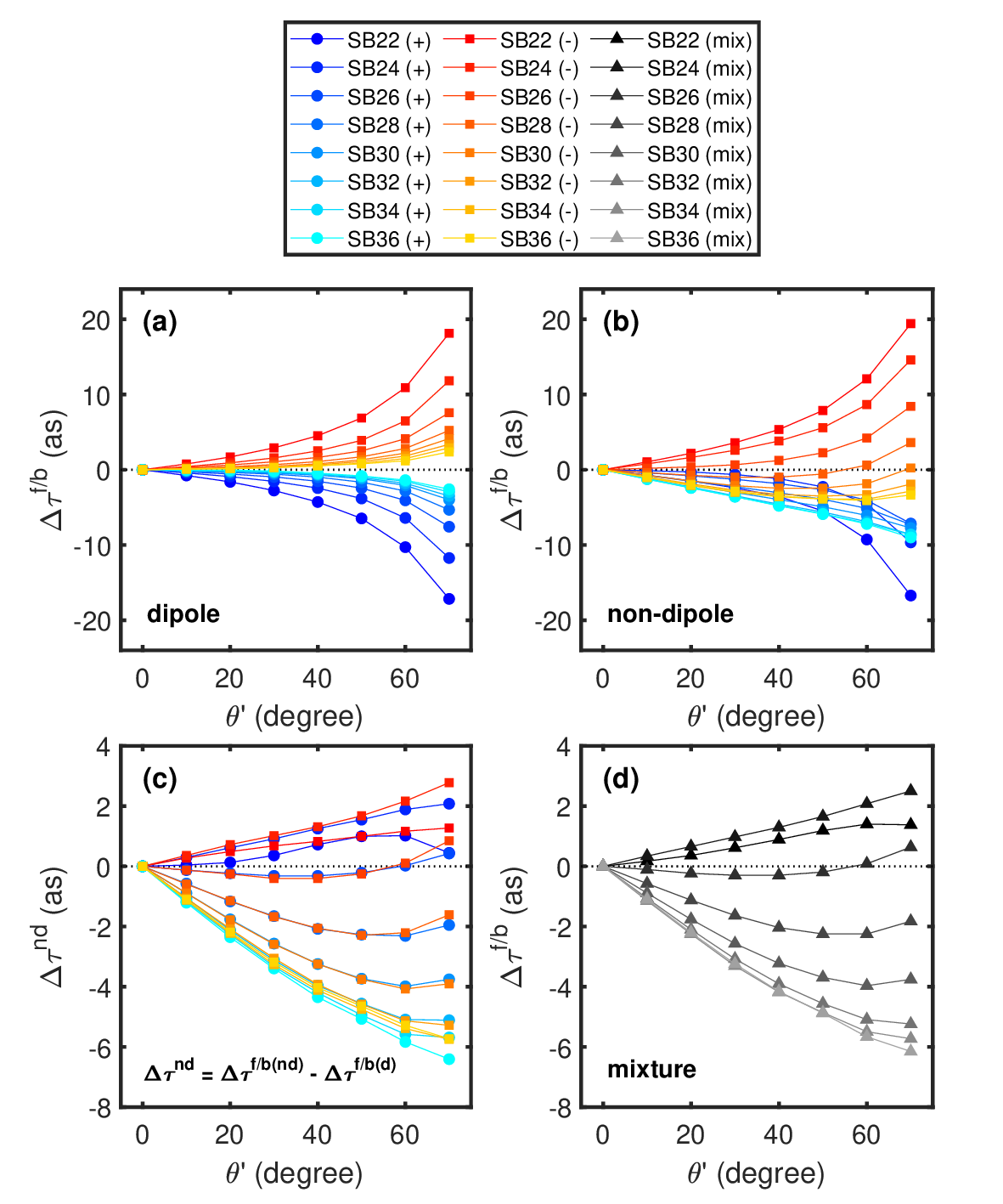}
    \caption{\label{F5}
        Forward-backward differential time delay for both enantiomers (a) in the dipole approximation and (b) including non-dipole effects. (c) Difference between the non-dipole and dipole results. (d) Forward-backward differential time delay for the racemic mixture. Here $\theta'=90^{\circ}-\theta$ is the emission angle relative to the polarization plane.
        }
\end{figure}

We evaluated $\Delta\mathrm{FBA}$ for the SBs, as shown in Figs.~\ref{F4}(k)--\ref{F4}(o). The strongest changes occur for $30^{\circ}<\theta<60^{\circ}$ and $120^{\circ}<\theta<150^{\circ}$, and the non-dipole contribution oscillates with the azimuthal angle $\varphi$. With increasing sideband energy, this oscillation becomes strong enough to induce local sign reversals. Nevertheless, the overall sign remains positive for $\theta<90^{\circ}$ and negative for $\theta>90^{\circ}$, again indicating preferential momentum transfer along the propagation direction. Figures~\ref{F4}(p)--\ref{F4}(t) show that the FBA calculated for a racemic mixture, which closely reproduces $\Delta\mathrm{FBA}$, confirming that the racemic subtraction protocol also applies to interferometric sidebands.

The sideband phase provides an additional observable. In chiral photoionization, forward and backward electron wave packets can acquire different phases, which are measured as differential photoionization time delays. Within second-order perturbation theory, the sideband angular distribution can be written as~\cite{10.1117/1.AP.8.1.015001}
\begin{equation}
    \begin{aligned}
        I(k,\theta,\varphi)=A(k,\theta)-B(k,\theta)\cos[2\varphi-\phi(k,\theta)].
    \end{aligned}
    \label{E13}
\end{equation}
Here $\phi(k,\theta)$ is the RABBITT phase, and $\tau(k,\theta)=\phi(k,\theta)/2\omega$ is the photoionization time delay for IR angular frequency $\omega$. We characterize the phase asymmetry by the forward-backward differential delay~\cite{doi:10.1126/science.aao5624,10.1117/1.AP.8.1.015001}, $\Delta\tau^{\text{f/b}}=\tau^{\text{f}}-\tau^{\text{b}}$. In the dipole approximation [Fig.~\ref{F5}(a)], $\Delta\tau^{\text{f/b}}$ grows with the angle from the polarization plane and is larger for low-energy electrons. Its sign reverses between enantiomers, as expected for a chiral phase response. Including non-dipole terms [Fig.~\ref{F5}(b)] substantially modifies the differential delay, especially for high-energy sidebands. For enantiomer (+), the high-energy delay increases and can exceed the low-energy delay; for enantiomer (-), the delay changes sign beginning at sideband 28 and the reversal strengthens at higher energy.

We define the non-dipole delay shift as $\Delta\tau^{\text{nd}}=\Delta\tau^{\text{f/b(nd)}}-\Delta\tau^{\text{f/b(d)}}$. As shown in Fig.~\ref{F5}(c), this shift increases with the angle relative to the polarization plane, exhibiting positive values at low energies and negative values at high energies. The values of $\Delta\tau^{\text{nd}}$ are nearly identical for the two enantiomers, indicating that the non-dipole delay shift is also enantiomer independent. Consequently, a racemic mixture can calibrate the non-dipole delay contribution in the same way that it calibrates the yield asymmetry. Figure~\ref{F5}(d) confirms this expectation: the racemic differential delay agrees with $\Delta\tau^{\text{nd}}$. Although the predicted non-dipole delay is only a few attoseconds, it will become increasingly relevant as chiral attosecond interferometry approaches this precision.

\section{Conclusion}

In conclusion, we have shown that non-dipole interactions are an essential systematic contribution to circular RABBITT measurements of chiral molecules. In both one-photon main peaks and two-photon sidebands, photon-momentum transfer modifies the forward-backward photoelectron asymmetry, enhances high-energy emission along the propagation direction, and can reverse the sign of the measured FBA. The two-pathway interference underlying RABBITT amplifies these non-dipole effects, making them visible at lower electron energies than in single-photon ionization. A central result is that the leading non-dipole contribution is nearly enantiomer independent. This makes it possible to measure the non-dipole background with a racemic mixture and subtract it from enantiopure measurements to recover the purely chiral asymmetry. The same strategy applies to the few-attosecond forward-backward delay shifts induced by non-dipole terms. These findings provide an experimentally actionable route to separate photon-momentum effects from genuine chiral response, an important step toward precision attosecond metrology of chiral electron dynamics.

\bigskip
\textbf{Acknowledgments}\textemdash This work was supported by National Natural Science Foundation of China (Grants No. U25D8005, 12374264, 12204545, 12434010),  National Key Research and Development Program of China (Grant No. 2023YFA1406800), Basic Research Support Program of Huazhong University of Science and Technology (2024BRA002). The computing work in this paper is supported by the Public Service Platform of High Performance Computing provided by Network and Computing Center of HUST.

\bigskip
\textbf{Note added}\textemdash Details of the numerical methods and second-order perturbation analysis for chiral molecules are provided in the Supplementary Material.

\bibliographystyle{unsrt}
\bibliography{ref_JPCL}

\end{document}